\documentclass{elsart}
\usepackage{natbib}
\usepackage{graphicx}
\usepackage{amssymb}

\def\apj{\ifmmode ApJ \else ApJ \fi}
\def\apjl{\ifmmode  ApJ \else ApJ \fi}
\def\aap{\ifmmode A\&A \else A\&A\fi}
\def\araa{\ifmmode ARA\&A \else ARA\&A\fi}
\def\cjaa{\ifmmode ChJAA \else ChJAA\fi}
\def\mnras{\ifmmode MNRAS \else MNRAS \fi}
\def\nat{\ifmmode Nature \else Nature \fi}

\def\prl{\ifmmode Phys. Rev. Lett. \else Phys. Rev. Lett.\fi}
\def\prd{\ifmmode Phys. Rev. D. \else Phys. Rev. D.\fi}
\def\bibitem{\par \noindent \hangafter=1 \hangindent=0.7 true cm}
\begin{document}
\newcommand{\vol}[2]{$\,$\rm #1\rm , #2.}
\begin{frontmatter}
\title{Study on the Correlations between the Twin Kilohertz Quasi-periodic
Oscillations in Low-mass X-ray Binaries}
\author{H. X. Yin$^{1*}$, Y. H. Zhao$^{1}$}
\address{$^{1}$ National Astronomical Observatories,
Chinese Academy of Sciences, Beijing 100012, China\\
         {\tt Email:yhx@lamost.org}}
\begin{abstract}
The recently updated data of the  twin kilohertz quasi-periodic
oscillations (kHz QPOs) in the neutron star low-mass X-ray binaries
are analyzed.
The power-law fitting $\nu_{1}=a(\nu_{2}/1000)^{b}$  and  linear
fitting $\nu_{2}=A\nu_{1}+B$ are applied, individually, to the
data points of four Z sources (GX 17+2, GX 340+0, GX 5-1 and Sco
X-1) and four Atoll sources (4U 0614+09, 4U 1608-52, 4U 1636-53
and 4U 1728-34).
The $\chi^{2}$-tests show that the power-law correlation and linear
correlation both can fit data well.
Moreover, the comparisons between the data and  the theoretical
models for kHz QPOs are discussed.
\end{abstract}
\begin{keyword}
QPOs; Accretion disks; Neutron stars; X-rays binaries.
\end{keyword}
\end{frontmatter}
\setlength{\parindent}{.25in}
\section{Introduction}
The kilohertz quasi-periodic oscillations (kHz QPOs) were firstly
discovered  in Sco X-1, a  luminous Z source in neutron star (NS)
low-mass X-ray binaries (LMXBs) (e.g. van der Klis et al. 1996),
and now they have been detected in twenty more sources (e.g. van
der Klis 2000, 2006, for reviews).
Usually, these kHz QPOs appear in pairs, the upper kHz QPO
frequency ($\nu_{2}$, hereafter the upper-frequency) and the lower
kHz QPO frequency ($\nu_{1}$, hereafter the lower-frequency),
which are discovered in  three classes of sources,  i.e. accretion
powered millisecond pulsars, bright Z sources and less luminous
Atoll sources (e.g., Hasinger \& van der Klis 1989).

%

%
The kHz QPO peak separation, $\Delta\nu=\nu_{2}-\nu_{1}$, in a given
source generally decreases with frequency, except the recently
detected kHz QPOs in Cir X-1, in which the peak separation increases
with frequency (Boutloukos et al. 2006). In addition, the variable
peak separations are not equal to the NS spin frequencies.
However, the averaged peak separation is found to be either close to
the spin frequency  or to half of it (e.g., van der Klis 2006;
Linares et al. 2005).

The above observations offer strong evidence against the simple
beat-frequency model, in which the lower-frequency is the beat
between the upper-frequency $\nu_{2}$ and the NS spin frequency
$\nu_{s}$ (e.g. Strohmayer et al. 1996; Zhang et al. 1997; Miller
et al. 1998), i.e. $\nu_{1}=\nu_{2}-\nu_{s}$.
Furthermore, with the discovery of  pairs of 30--450 Hz QPOs from a
few black-hole candidates with the frequency ratios 3:2 (e.g., van
der Klis 2006), Abramowicz et al.  (2003) reported that the ratios
of twin  kHz QPOs in  Sco X-1 tend to cluster around a value about
3:2, and they   argued  this fact  to be a promising link with the
black hole high-frequency QPOs (e.g. van der Klis 2006).

For the all Z and Atoll sources, the data plots of the
upper-frequency versus the lower-frequency can be fitted  by a power
law function (e.g., Zhang et al. 2006a), and also roughly fitted by
a linear function (Belloni et al. 2005). However, for the individual
kHz QPO source, for instance Sco X-1, its kHz QPOs can be well
fitted by a power law function (e.g. Psaltis et al. 1998; Yin et al.
2005).

In this paper, to investigate the twin kHz QPO correlation for the
individual Z or Atoll  source,  we fitted the data with a power-law
and a linear function for four typical Z sources and four typical
Atoll sources, and a comparison of both fittings by $\chi^{2}$-tests
is discussed in section 2, where comparisons with the models are
discussed.  The conclusions and consequences are given  in section
3.
\section{Correlations between twin kHz QPOs}
Until now, twin kHz QPOs have  been detected in 21 LMXBs, including
2 accretion powered millisecond X-ray pulsars, 8 Z sources and 11
Atoll sources, as listed in Tab. 1. In Fig. 1 and Fig. 2, we plotted
twin kHz QPO data for the Z sources and Atoll sources, showing  the
correlations of $\nu_{1}$ vs. $\nu_{2}$, $\Delta\nu$ vs. $\nu_{2}$
and $\nu_{2}/\nu_{1}$ vs. $\nu_{2}$, where the power-law and linear
fitting lines for the eight Z and  Atoll sources  are presented. The
results of the fittings and $\chi^{2}~\mbox{-tests}$ are listed in
Tab. 2.

\subsection{A  power law fitting}
The power-law function is chosen as
\begin{equation}
 \nu_{1}= a
 \left(\frac{\nu_{2}}{1000~\mbox{Hz}}\right)^{b}~\mbox{Hz}
 \label{nu12}
 \end{equation}
to fit twin kHz QPO data points of all Atoll (Z) sources, as well as
4 individual Atoll (Z) sources,  separately. It is noted that a same
function was applied to the fitting of kHz QPOs of Sco X-1 by
Psaltis et al. (1998)  with a smaller set of kHz QPO data points.
The fitting results of the normalization coefficient $a$, the
power-law index $b$ and $\chi^{2}/d.o.f.$ for various cases are
listed in Tab. 2, which correspond to the fitting curves as
presented in Fig. 1.
We find that the power-law index for the fitting of all Z sources
(see Tab. 2) is 1.87,  obviously bigger than that of the fitting for
all Atoll sources (1.61). Then, for the individual case, the
power-law index for Z source  is generally bigger than that in Atoll
source, except GX 17+2.

\subsection{A linear  fitting}
For the same data sets, the linear fitting function is chosen as,
\begin{equation}
\nu_{2}=A\nu_{1}+B~\mbox{Hz} \label{lin}\;,
\end{equation}
which was exploited  by Belloni et al. (2005) to discuss the kHz QPO
fitting in Sco X-1, 4U 1608-52, 4U 1636-53, 4U 1728-34 and 4U
1820-30.
%
By means of the $\chi^{2}~\mbox{-tests}$,  as shown in Tab. 2, we
find that the linear fitting concordes with the data well in some
cases, and there is no much systematic difference between the linear
slope parameters of the Atoll sources and  those of  Z sources.

\subsection{Comparison between the power-law  and the linear correlation}
As a comparison between models and the data,  it is remarked that
the relativistic precession model (e.g. Stella \& Vietri 1999) and
the Alfv\'en wave oscillation model (e.g. Zhang 2004; Li \& Zhang
2005) both can lead to power-law relations approximately, and then
the beat-frequency model (e.g. Miller et al. 1998) and the 3:2
resonance model (e.g. Abramowicz et al. 2003, this model is
successfully applied to black hole candidates) predicted the linear
relations between twin kHz QPO frequencies in the lowest
approximation (Abramowicz et al. 2005).
In Tab. 2, we can see that the $\chi^{2}/d.o.f.$ of the power-law
relation is usually less than the linear one for the same source,
except the two Atoll sources 4U 0614+09 and 4U 1636-53. And a linear
function cannot give a firstly increasing and then decreasing
tendency of all Z data as shown in Fig. 2b. But a power-law one
would fit it well as shown in Fig. 1b. So, these maybe mean that a
power-law correlation is better than a linear one.

\subsection{Testing the constant peak separation $\Delta\nu=300~\mbox{Hz}$}
Since the discovery of kHz QPOs,  it is known that the peak
separation   for Sco X-1 (van der Klis et al 1997; M\'endez \& van
der Klis 2000) is a  not constant, and the same is true for the
other Z sources, e.g., GX 17+2 (Homan et al. 2002) and Cir X-1
(Boutloukos 2006).
As for the  Atoll sources,   the  peak separation of 4U~1728--34
(Migliari, van der Klis, \& Fender 2003; M\'endez \& van der Klis
1999)  is  always significantly lower than the burst oscillation
 frequency, and  the peak separation of
  4U~1636--53 (Jonker, M\'endez, \& van der Klis 2002b; M\'endez, van
der Klis, \& van Paradijs 1998)  is   varying between being lower
and higher than half the spin frequency. In addition, 4U 1608-52 (
M\'endez et al. 1998) and 4U 1735-44 (Ford et al. 1998) are found to
share the varied peak separations.

In Fig. 1b or Fig. 2b, we show that the peak separations in all Z
sources
decrease (increase) systematically with the upper frequency if the
upper frequency is larger (less) than $\sim$700 Hz (e.g. van der
Klis 2000, 2006; Boutloukos et al. 2006). But this firstly
increasing and then decreasing with frequency is not clearly found
for the kHz QPO data of all Atoll sources, as shown in Fig. 1e and
Fig. 2e, which perhaps is on account of the less amount of data in
the low kHz QPO frequencies in Atoll sources.
From Fig. 1 and Fig. 2,  we find  that the peak separations are
scattered in a  wide range of frequency for each source. Therefore,
a constant peak separation, i.e. $\Delta\nu = 300~\mbox{Hz}$, cannot
fit for these data.

Fig. 3 shows the results of $\chi^{2}$-tests against a general
constant peak separation of the twin kHz QPOs in the 8 individual
sources. The minimum $\chi^{2}/d.o.f.$ are all with values $>>$ 1 ,
which means that any constant peak separation model cannot fit for
these data anymore.

\subsection{Testing the constant peak ratio $\nu_{2}/\nu_{1}=3/2$}

From Fig. 1c (Fig. 1f) or Fig. 2c (Fig. 2f), we find that twin
frequency ratios distribute in a wide range from 1.2 to 4.2, with
the averaged value 1.73 (1.50) for all known Z (Atoll) data.
Obviously, a constant ratio $\nu_{2}/\nu_{1}=3/2$, which can be
applicable to some  black hole QPO sources, is not consistent with
the observed NS/LMXB data.
In the $\nu_{2}/\nu_{1} ~\mbox{vs.} ~\nu_{2}$ plots of Fig. 1 and
Fig. 2, the frequency ratios systematically decrease  with the
upper-frequency for both Z and Atoll sources.
In detail, the incompatible 3:2 ratio peak distribution has been
also studied by Belloni et al. (2005) in several sources, who
showed that the distribution of QPO frequencies in {Sco X-1}, {4U
1608--52}, {4U 1636--53},  {4U 1728--34}, and {4U 1820--30} is
multi-peaked, with the peaks occurring at the different $\nu_{2} /
\nu_{1}$ ratios.
\section{Conclusions}
In this paper, the updated  data sets  of twin kHz QPO frequencies
simultaneously detected in NS LMXBs are analyzed, and  the
power-law and linear fittings are studied  for the individual
Z/Atoll and all Z/Atoll sources, respectively. Our main
conclusions are presented as follows.
(1)In Fig.1 and Fig.2, we can notice that a simple constant peak
separation model, such as the beat-frequency model (e.g. Strohmayer
et al. 1996; Zhang et al. 1997; Miller et al. 1998), or a constant
peak ratio assumption, as in a naive extrapolation of the observed
resonant frequency ratio from the black hole sources to the neutron
stars (e.g. Abramowicz et al. 2003), cannot fit the observed data.
Namely, any simple constant peak separation and constant peak ratio
models are generally inconsistent with the data.
The peak separations in all Z sources tend to increase (decrease)
with the upper-frequency if the upper-frequency is less (larger)
than $\sim$700 Hz.  But this tendency does not appear in all Atoll
sources because of less amount of data at low kHz QPO frequency.
Statistically, the twin frequency ratios tend to decrease with the
upper-frequency in both Z and Atoll sources.
(2)Our results show that the  index  of the fitted power-law
relation of Z source is generally  bigger than that  of Atoll
source, except GX 17+2. On the consideration of model, this
different index value in Z and Atoll sources might be related to the
diversity in their luminosity or magnetic field.
However the linear correlations do not show any systematic
differences between Z and Atoll sources.
(3)The power-law fitting is somewhat better than the linear one for
most of the sources, because the $\chi^{2}/d.o.f.$ value  of the
power-law correlation is generally less  than that of linear one,
and a linear correlation cannot give the firstly increasing and then
decreasing tendency of peak separations in all Z data. As a
comparison with model's prediction, we mention the Relativistic
Precession model (e.g. Stella \& Vietri 1999) and the Alfv\'en Wave
Oscillation model (Zhang 2004), since both models can give an
approximated power-law correlation between the twin kHz QPOs,
however none of them can distinguish the influences of the
luminosity of Z and Atoll sources.

As a summary,  if the future data still support the conclusions
obtained in the paper, they will pose the meaningful constraints on
the models for explaining kHz QPOs.

\section{Acknowledgements} We are grateful for T. Belloni, M.
M\'endez, D. Psaltis and J. Homan for providing the QPO data, and
thank C.M. Zhang for discussions. We highly appreciate the anonymous
reviewers for their helpful comments.
\section{References}
\bibitem{}
Abramowicz, M.A., Bulik, T., Bursa, M., et al. Evidence for a 2:3
resonance in Sco X-1 kHz QPOs. \aap\vol{404}{L21--L24} 2003.

\bibitem{}
Abramowicz, M.A., Barret, D., Bursa, M., et al. AN,
\vol{326}{864--866} 2005.

\bibitem{}
Belloni,T., Psaltis, D.,  \& van der Klis, M.  A Unified
Description of the Timing Features of Accreting X-Ray Binaries.
\apj\vol{572}{392--406} 2002.

\bibitem{}
Belloni, T., M\'endez, M., \& Homan, J. The distribution of kHz
QPO frequencies in bright low mass X-ray binaries.
\aap\vol{437}{209--216} 2005.

\bibitem{}
Boutloukos, S., van der Klis, M., Altamirano, D., et al. Discovery
of twin kHz QPOs in the peculiar X-ray binary Circinus X-1. \apj
in press 2006. (astro-ph/0608089)

\bibitem{}
Di Salvo, T., M\'endez, M., \& van der Klis, M. On the correlated
spectral and timing properties of 4U 1636-53: An atoll source at
high accretion rates. \aap\vol{406}{177--192} 2003.

\bibitem{}
Hasinger, G., \& van der Klis, M. Two patterns of correlated X-ray
timing and spectral behaviour in low-mass X-ray
binaries.\aap\vol{225}{79--96} 1989.

\bibitem{}
Homan, J., van der Klis, M., Jonker, P.G., et al. RXTE
Observations of the Neutron Star Low-Mass X-Ray Binary GX 17+2:
Correlated X-Ray Spectral and Timing Behavior.
\apj\vol{568}{878--900} 2002.

\bibitem{}
Jonker, P.G., van der Klis, M., Wijnands, et al. The Power
Spectral Properties of the Z Source GX 340+0.
\apj\vol{537}{374--386} 2000.

\bibitem{}
Jonker, P.G., van der Klis, M., Homan, J., et al. Low- and
high-frequency variability as a function of spectral properties in
the bright X-ray binary GX 5-1. \mnras\vol{333}{665--678} 2002a.

\bibitem{}
Jonker, P.G., M\'endez, M., \& van der Klis, M. Kilohertz
quasi-periodic oscillations difference frequency exceeds inferred
spin frequency in 4U 1636-53. \mnras\vol{336}{L1--L5} 2002b.

\bibitem{}
Li, X.D., \& Zhang, C.M. A Model for Twin Kilohertz Quasi-periodic
Oscillations in Neutron Star Low-Mass X-Ray Binaries.
\apj\vol{635}{L57--L60} 2005.

\bibitem{}
Linares, M., van der Klis, M., Altamilano, D. et al. Discovery of
Kilohertz Quasi-periodic Oscillations and Shifted Frequency
Correlations in the Accreting Millisecond Pulsar XTE J1807-294.
\apj\vol{634}{1250--1260} 2005.

\bibitem{}
Markwardt, C.B., Strohmayer, T.E., \& Swank, J.H. Observation of
Kilohertz Quasi-periodic Oscillations from the Atoll Source 4U
1702-429 by the Rossi X-Ray Timing Explorer.
\apjl\vol{512}{L125--L129} 1999.

\bibitem{}
M\'endez, M., van der Klis, M., \& van Paradijs, J. Difference
Frequency of Kilohertz QPOs Not Equal to Half the Burst
Oscillation Frequency in 4U 1636-53. \apjl\vol{506}{L117--L119}
1998.

\bibitem{}
M\'endez, M., van der Klis, M., Wijnands, R., et al. Kilohertz
Quasi-periodic Oscillation Peak Separation Is Not Constant in the
Atoll Source 4U 1608-52. \apjl\vol{505}{L23--L26} 1998.

\bibitem{}
M\'endez, M., \& van der Klis, M. Precise Measurements of the
Kilohertz Quasi-periodic Oscillations in 4U 1728-34.
\apj\vol{517}{L5--L54} 1999.

\bibitem{}
M\'endez, M., van der Klis, M. The harmonic and sideband structure
of the kilohertz quasi-periodic oscillations in Sco X-1.
\mnras\vol{318}{938--942} 2000.

\bibitem{}
Migliari, S., van der Klis, M., \& Fender, R. Evidence of a
decrease of kHz quasi-periodic oscillation peak separation towards
low frequencies in 4U 1728-34 (GX 354-0).
\mnras\vol{345}{L35--L39} 2003.

\bibitem{}
Miller, M.C., Lamb, F.K., \& Psaltis, D. . Sonic-Point Model of
Kilohertz Quasi-periodic Brightness Oscillations in Low-Mass X-Ray
Binaries.\apj\vol{508}{791--830} 1998.

\bibitem{}
O'Neill, P.M., Kuulkers, E., Sood, R.~K., et al. The X-ray
fast-time variability of Sco X-2 (GX 349+2) with RXTE.
\mnras\vol{336}{217--232} 2002.


\bibitem{}
Psaltis, D., M\'endez, M., Wijnands, R., et al. The Beat-Frequency
Interpretation of Kilohertz Quasi-periodic Oscillations in Neutron
Star Low-Mass X-Ray Binaries. \apj\vol{501}{L95--L99} 1998.

\bibitem{}
Psaltis, D., Wijnands, R., Homan, J., Jonker, et al. On the
Magnetospheric Beat-Frequency and Lense-Thirring Interpretations
of the Horizontal-Branch Oscillation in the Z Sources.
\apj\vol{520}{763--775} 1999a.

\bibitem{}
Psaltis, D., Belloni, T. \& van der Klis, M. Correlations in
Quasi-periodic Oscillation and Noise Frequencies among Neutron
Star and Black Hole X-Ray Binaries. \apj\vol{520}{262--270} 1999b.

\bibitem{}
Stella, L., Vietri, M. \& Morsink, S. Correlations in the
Quasi-periodic Oscillation Frequencies of Low-Mass X-Ray Binaries
and the Relativistic Precession Model. \apj\vol{524}{L63--L66}
1999.

\bibitem{}
Strohmayer, T., Zhang, W., Smale, A., et al. Millisecond X-Ray
Variability from an Accreting Neutron Star System.
\apj\vol{469}{L9--L12} 1996.

\bibitem{}
van der Klis, M., Swank, J.H., Zhang, W., et al. Discovery of
Submillisecond Quasi-periodic Oscillations in the X-Ray Flux of
Scorpius X-1. \apj\vol{469}{L1--L4} 1996.

\bibitem{}
van der Klis, M., Wijnands, R., Horne, D. et al. Kilohertz
Quasi-Periodic Oscillation Peak Separation Is Not Constant in
Scorpius X-1. \apj\vol{481}{L97--L100} 1997.

\bibitem{}
van der Klis, M. Millisecond Oscillations in X-ray Binaries.
\araa\vol{38}{717--760} 2000.

\bibitem{}
van der Klis, M. Rapid X-Ray Variability. in Compact stellar X-ray
sources, W.H.G. Lewin \& M. van der Klis (eds.), Cambridge
University Press, p.39. 2006. (astro-ph/0410551)

\bibitem{}
van Straaten, S., Ford, E.C., van der Klis, M., et al. Relations
between Timing Features and Colors in the X-Ray Binary 4U 0614+09.
\apj\vol{540}{1049--1061} 2000.

\bibitem{}
van Straaten, S., van der Klis, M., Di Salvo, T., et al. A
Multi-Lorentzian Timing Study of the Atoll Sources 4U 0614+09 and
4U 1728-34. \apj\vol{568}{912--930} 2002.

\bibitem{}
van Straaten, S., van der Klis, M., \& M\' endez, M. The Atoll
Source States of 4U 1608-52. \apj\vol{596}{1155--1176} 2003.

\bibitem{}
Wijnands, R.,  van der Klis, M.,  Homan, J., et al. Quasi-periodic
X-ray brightness fluctuations in an accreting millisecond pulsar.
\nat\vol{424}{44--47} 2003.

\bibitem{}
Yin, H.X., Zhang, C.M., Zhao, Y.H., et al. A Study on the
Correlations between the Twin kHz QPO frequencies in Sco X-1.
\cjaa\vol{5}{595--600} 2005.

\bibitem{}
Zhang, C.M. The MHD Alfven wave oscillation model of kHz Quasi
Periodic Oscillations of Accreting X-ray binaries.
\aap\vol{423}{401--404} 2004.

\bibitem{}
Zhang, C.M., Yin, H.X., Zhao, Y.H., et al. The correlations
between the twin kHz quasi-periodic oscillation frequencies of
low-mass X-ray binaries. \mnras\vol{366}{1373--1377} 2006a.

\bibitem{}
Zhang, F., Qu, J.L., Zhang, C.M., et al. Timing Features of the
Accretion-driven Millisecond X-Ray Pulsar XTE J1807-294 in the
2003 March Outburst. \apj\vol{646}{1116--1124} 2006b.

\bibitem{}
Zhang, W., Strohmayer, T.E., \& Swank, J.H. Neutron Star Masses
and Radii as Inferred from Kilohertz Quasi-periodic Oscillations.
\apj\vol{482}{L167--L170} 1997.

\begin{figure*}
\begin{tabular}{cc}
  \includegraphics[height=15.0cm,width=0.5\hsize]{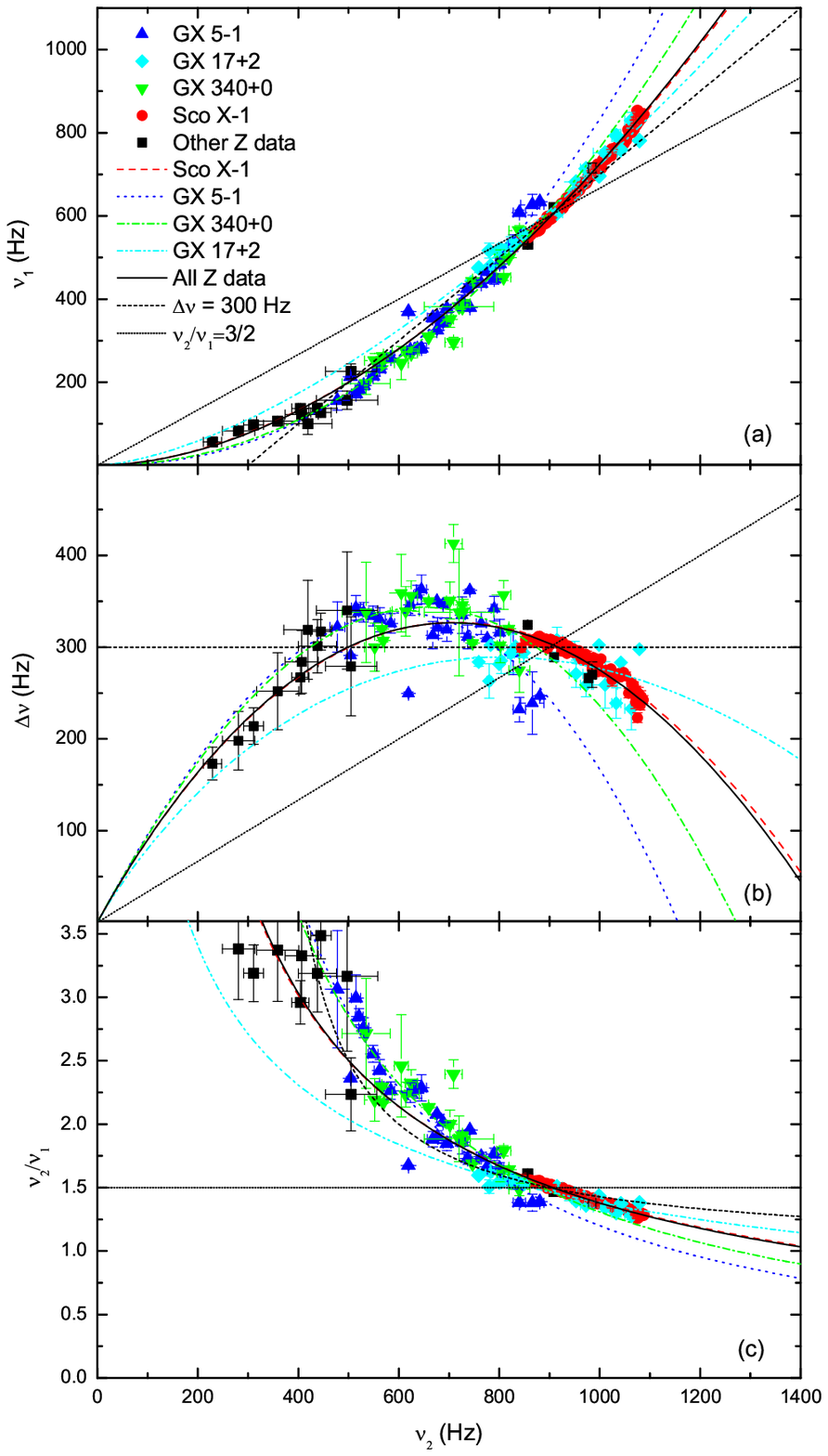}
  \includegraphics[height=15.0cm,width=0.5\hsize]{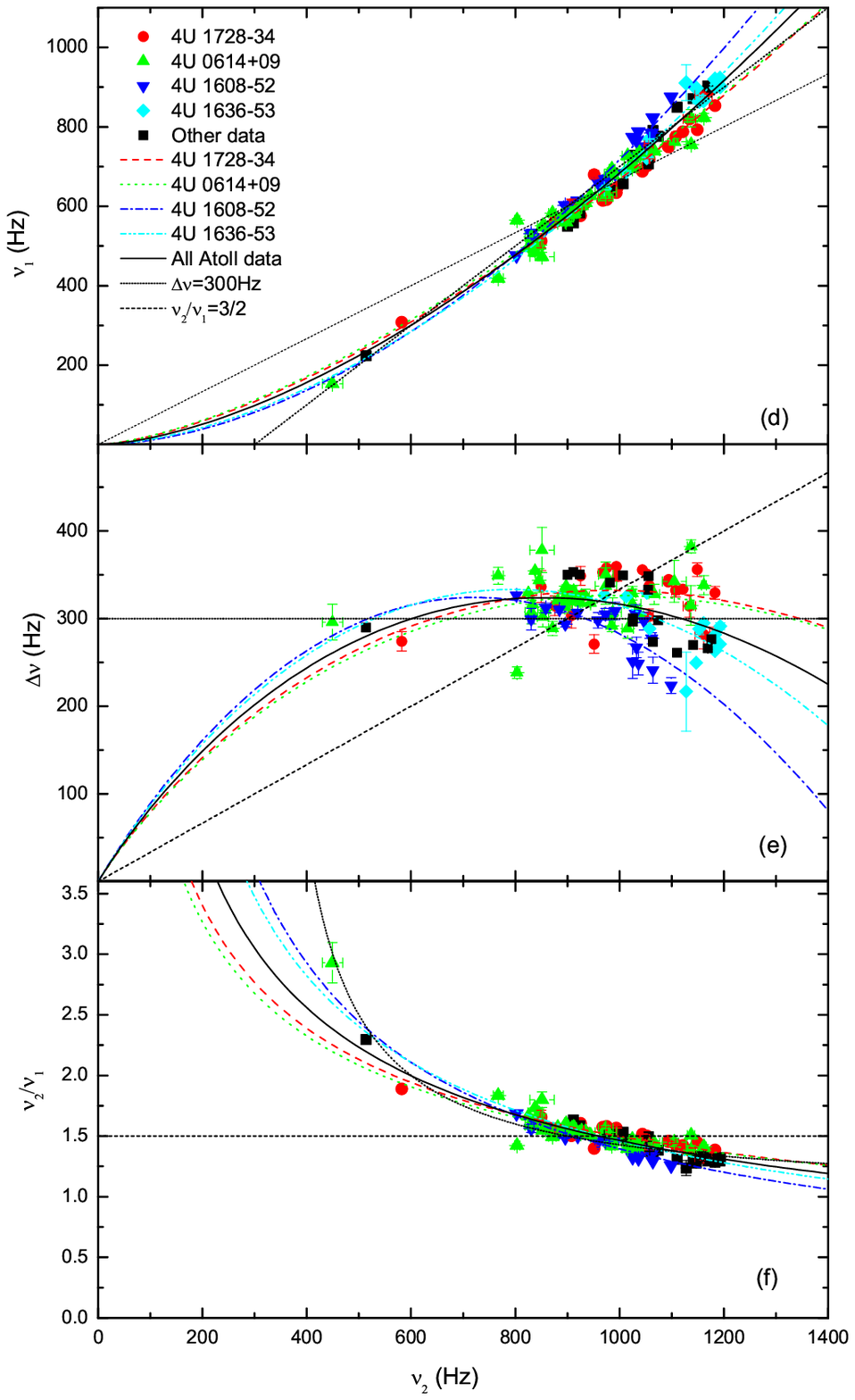}
  \end{tabular}
  \caption{\bf Plots of a and d $\nu_{1}$ vs. $\nu_{2}$, b and e $\Delta\nu$ vs.
$\nu_{2}$ and c and f $\nu_{2}/\nu_{1}$ vs. $\nu_{2}$ for  Z
sources and Atoll sources. Power-law fitting lines and the two
reference lines ($\nu_{2}/\nu_{1}=3/2$, and
$\Delta\nu=300~\mbox{Hz}$) are presented also. }
  \label{Fig:fig1}
\end{figure*}

\begin{figure*}
\begin{tabular}{cc}
  \includegraphics[height=15.0cm,width=0.5\hsize]{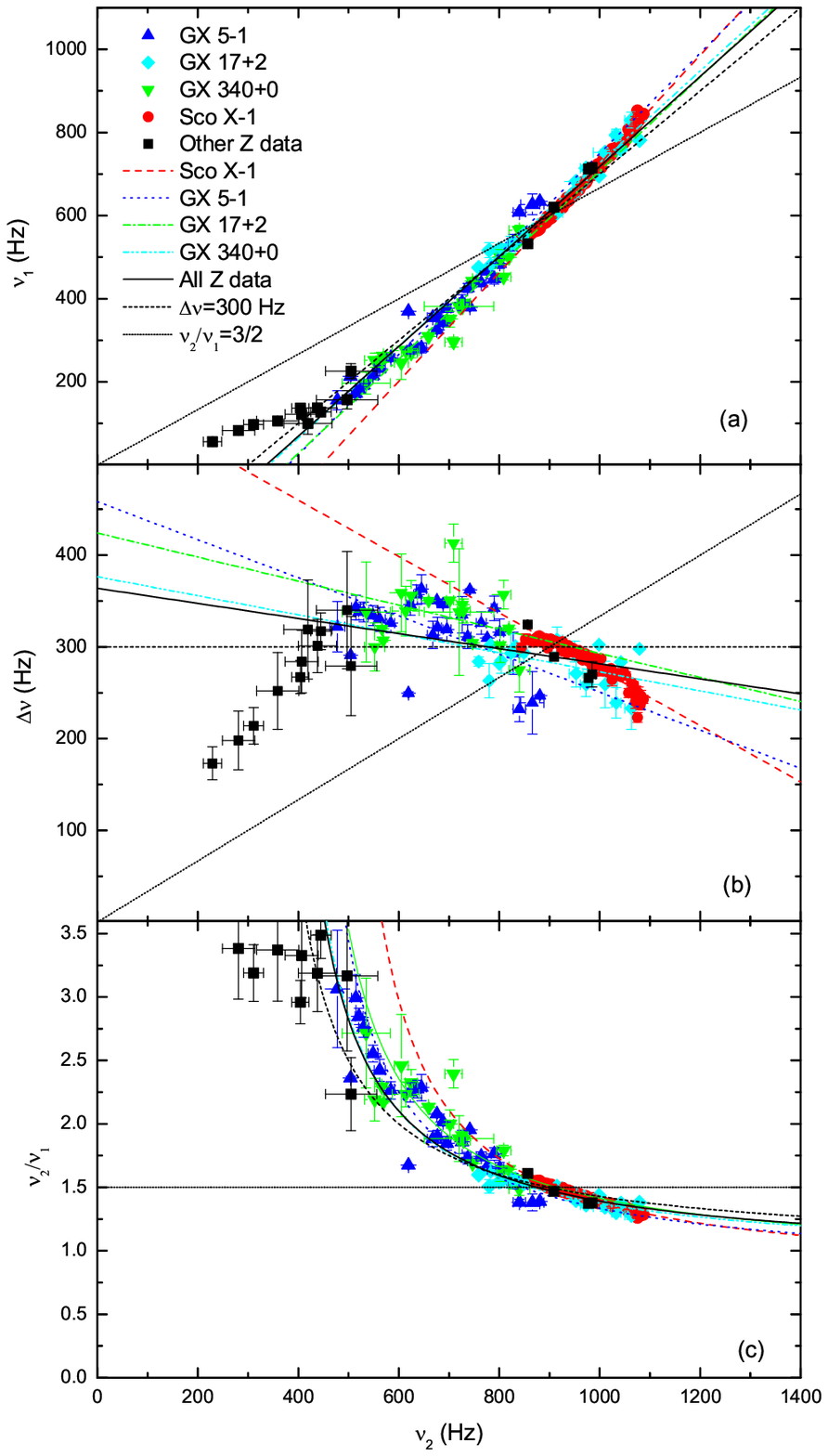}
  \includegraphics[height=15.0cm,width=0.5\hsize]{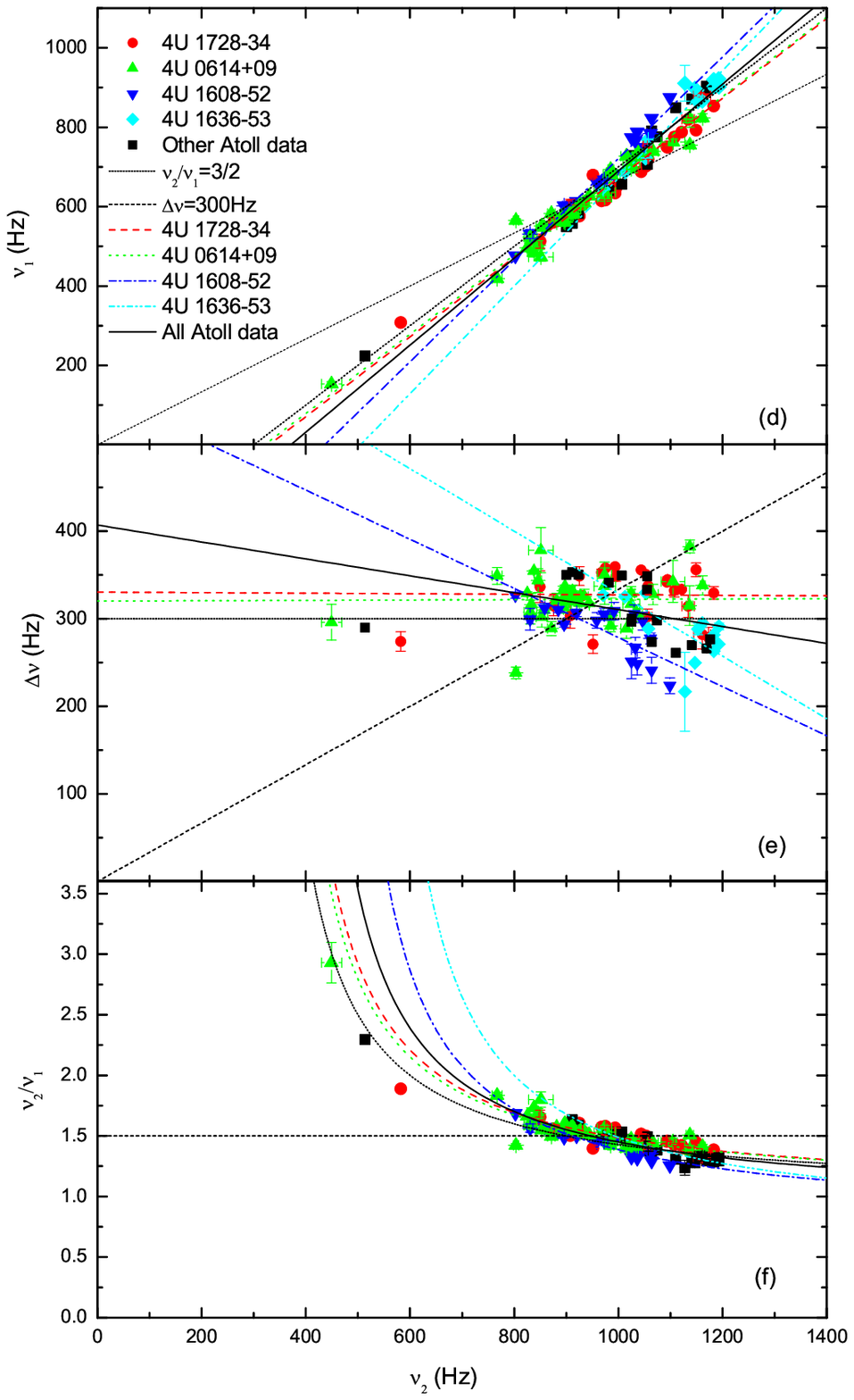}
  \end{tabular}
  \caption{\bf Plots of a and d $\nu_{1}$ vs. $\nu_{2}$,  b and e $\Delta\nu$ vs.
$\nu_{2}$ and  c and f $\nu_{2}/\nu_{1}$ vs. $\nu_{2}$ for Z
sources and Atoll sources. Linear fitting lines and the two
reference lines ($\nu_{2}/\nu_{1}=3/2$, and
$\Delta\nu=300~\mbox{Hz}$) are presented also.}
  \label{Fig:fig2}
\end{figure*}

\begin{figure}
\includegraphics[width=\hsize]{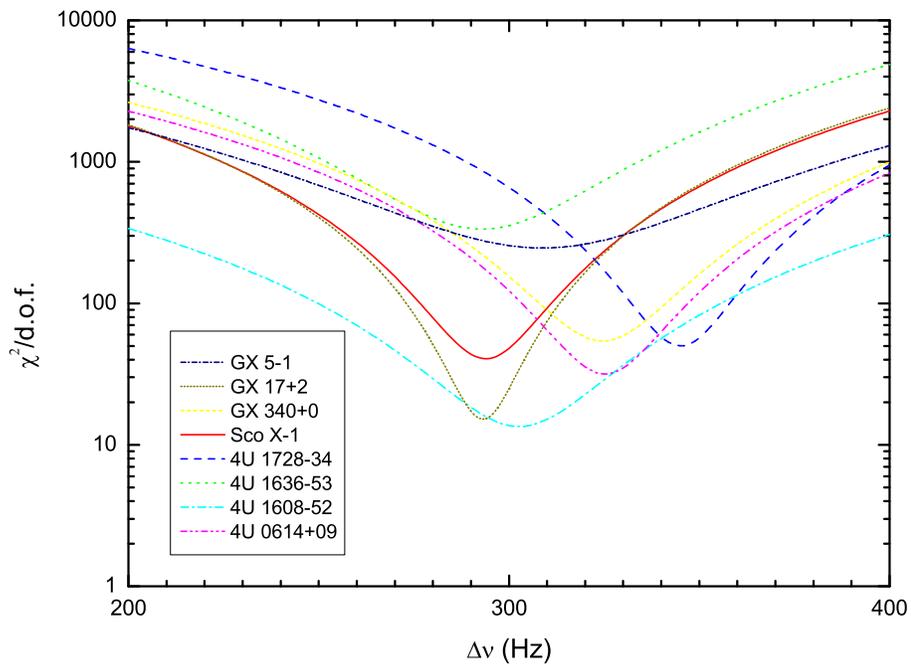}
\caption{\bf $\chi^{2}-$tests for a constant peak separation of the
8 individual sources list in Tab. 2.} \label{Fig:fig3}
\end{figure}

\clearpage
\newpage
\vskip 1.0cm
\begin{table}
\begin{minipage}{\linewidth}
  \caption{\bf List of LMXBs with the simultaneously detected twin kHz QPOs.}
  \label{list}
  \begin{center}
  \begin{tabular}{lccccl}
  \hline
  \hline
   Sources          & $\nu_{1}^{(1)}$   & $\nu_{2}^{(2)}$ & $\Delta\nu^{(3)}$ & $\nu_{2}/\nu_{1}^{(4)}$  & References   \\
       & (Hz) & (Hz) & (Hz) &  & \\
  \hline
{\bf Millisecond pulsar (2)}                 \\
  \hline
XTE J1807-294      &127-360 & 353-587 & 179-247 & 1.51-2.78       &      1,2,                    \\
SAX J1808.4-3658   & 499   &    694   &  195       &  1.39       &      3                     \\
 \hline
{\bf Z source (8)}                \\
  \hline
Cir X-1     & 56-226    & 229-505   & 173-340    &  2.23-4.19            &    4\\
Sco X-1     &  544-852  & 844-1086  &223-312     &  1.26-1.57            &   B,M,K                 \\
GX 340+0    &   197-565 & 535-840   &275-413     &  1.49-2.72            &   B,K,P,5               \\
XTE J1701-462 &  620    & 909       & 289        & 1.47                  &  6\\
GX 349+2    &   712-715 & 978-985   &266-270     &  1.37-1.38            &   B,K,7                \\
GX 5-1      &  156-634  & 478-880   &232-363     &  1.38-3.06            &   B,K,P,8                \\
GX 17+2     &  475-830  & 759-1078  &233-308     &  1.28-1.60            &   B,K,P,9                \\
Cyg X-2     &     532   &  856.6    & 324        &   1.61                &   B,K,P                  \\
  \hline
{\bf Atoll source (11)}               \\
  \hline
4U 0614+09  &  153-823  & 449-1162  & 238-382    &   1.38-2.93           &   B,K,P,10,11             \\
4U 1608-52  &  476-876  & 802-1099  &224-327     &  1.26-1.69            &   M,B,K,12                 \\
4U 1636-53  &  644-921  & 971-1192  &217-329     &  1.24-1.51            &   B,K,P,13,14              \\
4U 1702-43  &  722      &  1055     & 333        &     1.46              &   K,P,15                  \\
4U 1705-44  &  776      &  1074     & 298        &     1.38              &   B,K,P             \\
4U 1728-34  &  308-894  & 582-1183  &271-359     &  1.31-1.89            &   B,K,P,11,16       \\
KS 1731-260 &  903      &  1169     & 266        &    1.29               &   B,K,P          \\
4U 1735-44  & 640-728   & 982-1026  &296-341     &  1.41-1.53            &   B,K,P             \\
4U 1820-30  &  790      &  1064     & 273        &    1.35               &   B,K,P                  \\
4U 1915-05  & 224-707   & 514-1055  &290-353     &  1.49-2.3             &   B,K,P             \\
XTEJ2123-058& 849-871   &1110-1140  &261-270     &  1.31-1.31            &   B,K,P                  \\
  \hline
  \hline
  \end{tabular}
  \end{center}
\vskip 0.3cm
\begin{tabular}{p{\linewidth}}
{\bf $^{(1)}$:the range of $\nu_{1}$; $^{(2)}$: the range of
$\nu_{2}$; $^{(3)}$: the range of $\Delta\nu$; $^{(4)}$: the range
of $\nu_{2}/\nu_{1}$. K: van der Klis 2000, van der Klis 2006; M:
M\'endez et al. 1998ab, M\'endez \& van der Klis 1999, 2000; B:
Belloni et al. 2002, Belloni et al. 2005; P: Psaltis et al.
1999ab.  1: Linares 2005; 2: Zhang et al. 2006b; 3: Wijnands et
al. 2003; 4: Boutloukos et al. 2006; 5: Jonker et al. 2000; 6:
Homan 2006 (personal communication); 7: O'Neill et al. 2002; 8:
Jonker et al. 2002a; 9: Homan et al. 2002; 10: van Straaten et al.
2002; 11: van Straaten et al. 2000; 12: van Straaten et al. 2003;
13: Di Salvo et al. 2003; 14: Jonker et al. 2002b; 15: Markwardt
et al. 1999; 16: Migliari et al. 2003.}
\end{tabular}
\end{minipage}
\end{table}

\clearpage
\newpage
\begin{table}
  \caption{List of the results of fittings and $\chi$$^{2}$-tests.}
  \label{Tab:table-2}
  \begin{center}\begin{tabular}{lc||ccc||ccc}
  \hline
  \hline
   &&\multicolumn{3}{c||}{\bf $\nu_{1}=\mbox{a}(\nu_{2}/(1000~\mbox{Hz}))^{b}~\mbox{Hz}$}&\multicolumn{3}{c}{\bf $\nu_{2}=\mbox{A}\nu_{1}$+B Hz}\\
  \cline{3-8}
  \raisebox{2.3ex}[]{Source$^{*}$}&\raisebox{2.3ex}[]{}  & a  &  b   & $\chi^{2}$/d.o.f.  &A  & B &$\chi^{2}$/d.o.f.  \\
  \hline
  {\bf Z source}\\
  \hline
Sco X-1         &&    721.95  $\pm$ 0.69    &  1.85 $\pm$  0.01 & 33.9/87  &    0.765 $\pm$  0.007 & 445.84 $\pm$ 4.73 & 54.8/87 \\
GX 340+0        &&    763.85  $\pm$38.03    &  2.12 $\pm$  0.15 & 21.3/17  &    0.884 $\pm$  0.067 & 374.96 $\pm$24.41 & 27.2/17 \\
GX 5-1          &&    833.02  $\pm$26.08    &  2.26 $\pm$  0.10 & 29.7/27  &    0.828 $\pm$  0.039 & 379.54 $\pm$15.10 & 43.7/27 \\
GX 17+2         &&    723.40  $\pm$ 5.53    &  1.56 $\pm$  0.06 & 11.2/19  &    0.906 $\pm$  0.038 & 341.13 $\pm$24.27 & 12.0/19 \\
All Z           &&    725.38  $\pm$ 2.50    &  1.87 $\pm$  0.02 & 205.8/169&    0.924 $\pm$  0.012 & 336.22 $\pm$ 6.81 & 380.1/169\\
  \hline
  {\bf Atoll source} \\
  \hline
4U 0614+09       &&    673.74  $\pm$ 4.77    &  1.49 $\pm$  0.06 & 80.1/40  &    1.002 $\pm$  0.033 & 320.75 $\pm$20.37&68.4/40\\
4U 1608-52       &&    717.36  $\pm$ 4.89    &  1.81 $\pm$  0.08 & 7.8/16   &    0.781 $\pm$  0.036 & 436.60 $\pm$25.12&9.1/16\\
4U 1636-53       &&    685.16  $\pm$15.44    &  1.72 $\pm$  0.16 & 10.7/11  &    0.737 $\pm$  0.064 & 505.18 $\pm$53.51&10.1/11\\
4U 1728-34       &&    667.86  $\pm$ 5.59    &  1.51 $\pm$  0.07 & 20.9/23  &    0.997 $\pm$  0.046 & 329.33 $\pm$32.09&25.5/23\\
All Atoll        &&    683.48  $\pm$ 3.01    &  1.61 $\pm$  0.04 & 293.3/109&    0.912 $\pm$  0.020 & 371.08 $\pm$13.91&308.7/109\\
 \hline
 \hline
\end{tabular}\end{center}
\end{table}

\end{document}